\documentclass[twocolumn,showpacs, superscriptaddress,
amssymb, nobibnotes, notitlepage,nofootinbib aps, prd]{revtex4}

\usepackage{amsmath}
\usepackage{amssymb}
\usepackage{amsthm}
\usepackage[pdftex]{color}
\usepackage[pdftex,colorlinks,citecolor=blue,linkcolor=blue,urlcolor=blue]{hyperref} 
\usepackage{graphicx}
\usepackage{dcolumn} 
\usepackage{bm} 

\usepackage{longtable}

\usepackage{ulem}   
\usepackage{comment} 
\usepackage{datetime}

\usepackage{tabularx}
\usepackage{float}
\restylefloat{table}

\newcommand{\beq}{\begin{equation}}
\newcommand{\eeq}{\end{equation}}
\newcommand{\bea}{\begin{eqnarray}}
\newcommand{\eea}{\end{eqnarray}}
\newcommand{\barr}{\begin{array}}
\newcommand{\earr}{\end{array}}

\long\def\begincomment#1\endcomment{}

\newcommand{\g}{\gamma}

\newtheorem{remark}{Remark}

\begin{document}


\title{ \Large Noncommutative Dirac and Klein-Gordon oscillators in the background of cosmic string:  spectrum and dynamics
}

\author{Baloitcha Ezinvi} \email{ezinvi.baloitcha@cipma.uac.bj}  
\affiliation{Facult\'e des Sciences et Techniques,\\ International Chair in Mathematical Physics and Applications (ICMPA-UNESCO Chair), University of Abomey-Calavi, 072B.P.50, Cotonou, Republic of Benin}

\author{Mahouton Norbert Hounkonnou} \email{norbert.hounkonnou@cipma.uac.bj}
\affiliation{Facult\'e des Sciences et Techniques,\\ International Chair in Mathematical Physics and Applications (ICMPA-UNESCO Chair), University of Abomey-Calavi, 072B.P.50, Cotonou, Republic of Benin}

\author{Emanonfi Elias N'Dolo}
\email{emanonfieliasndolo@yahoo.fr}
\affiliation{Facult\'e des Sciences et Techniques,\\ International Chair in Mathematical Physics and Applications (ICMPA-UNESCO Chair), University of Abomey-Calavi, 072B.P.50, Cotonou, Republic of Benin}

\author{Dine Ousmane Samary}
\email{dine.ousmane.samary@aei.mpg.de}

 \affiliation{Max Planck Institute for Gravitational Physics, Albert Einstein Institute, Am M\"uhlenberg 1, 14476, Potsdam, Germany}
\affiliation{Facult\'e des Sciences et Techniques,\\ International Chair in Mathematical Physics and Applications (ICMPA-UNESCO Chair), University of Abomey-Calavi, 072B.P.50, Cotonou, Republic of Benin}

\date{\today}

\begin{abstract}
From a study of an oscillator in a $4D$ NC spacetime, we establish the Hamilton equations of motion. The formers are solved to give the oscillator position and momentum coordinates. These coordinates are used to build a metric similar to that describing a cosmic string. On this basis, Dirac and Klein-Gordon oscillators are investigated. Their spectrum and dynamics are analysed giving rise to novel interesting properties. 
\end{abstract}


\maketitle

\section{Introduction} 
Noncommutative (NC) field theory could play a special role in the description of particle physics  near the Planck length $\lambda_p=\sqrt{G\hbar/c^3}$. This theory  was the subject of intense study during the last three decades and provided a very interesting new class of quantum field theories with intriguing and sometimes
unexpected features. The idea of noncommutativity of spacetime came from Snyder \cite{Snyder:1946qz}. Its geometric analysis  was given  by Alain Connes \cite{Connes:1994yd}-\cite{Connes:1990qp}. The wide class of works on this subject and  the physical implications such as the
quantum Hall effect \cite{Harms:2006dv}-\cite{Scholtz:2005vg}, the string theory static solutions \cite{Seiberg:1999vs}, the matrix model or the $2D$ quantum gravitation theory \cite{Zinn-Justin:2014wva}-\cite{Livine:2009zz}   opened new outlook on  the study of physics.
 This made them particularly interesting and challenging for purposes of particle physics model building.


However, the NC spacetime generalizes the  ordinary space by assuming  the nonvanishing commutation relations  between coordinates  as $[\hat x^\mu,\hat x^\nu]=i\theta^{\mu\nu}$, where $(\theta^{\mu\nu})$ is skew-symmetric constant tensor. 
The operator algebra of such NC spacetime can be represented by the algebra of functions when the ordinary multiplication of functions is replaced by the so-called Moyal star-product: 
\bea
(f\star g)(x)={\rm \bf m}\Big[\exp\Big[\theta^{\mu\nu}\partial_\mu\otimes \partial_\nu\Big] (f\otimes g)(x)\Big],\,\,\cr
{\rm \bf m}(f\otimes g)=f\cdot g,\,\,\, f,g\in C^{\infty}(\mathbb{R}^D)
\eea
(see \cite{Szabo:2001kg} and references therein).  Another noncommutativity is described when the momentum components become NC, i.e. $[\hat p_\mu,\hat p_\nu]=i\bar{\theta}_{\mu\nu}$, where $(\bar\theta_{\mu\nu})$ is related to the magnetic field \cite{Gouba:2016iar}.  Several physical models were studied in this NC spacetime, such as the model of  harmonic oscillator, the  dynamics of  the relativistic particles, and the scattering theory.

Recently, the Dirac  and KG oscillators were studied  in curve  spacetime  by  introducing the tedrad fields $e_a^\mu,$ or, equivalently, the metric ${\rm g}^{\mu\nu}=e_a^\mu e_b^\nu \eta^{ab},$ where $\eta$ is the flat spacetime metric \cite{ito}-\cite{Akcay:2016wgl}. 
These models were also implemented in the topologycal defect background metric \cite{Carvalho:2011krd}.    An important question that we address here in this paper  is the   effects of noncommutativity on the dynamics of the spin orbit particle.  We show that the noncommutativity of the spacetime  transforms the Minkowski metric to  the so-called  cosmic string background. 
 By giving the solution of the oscillator dynamics  using the Hamilton equations of motion, we derive the corresponding deformation of the spacetime metric, which depends on the parameter $\theta$ and is similar to that describing a cosmic string.
 As application,  we are  interested  in relativistic particles described by  Dirac and Klein-Gordon (KG) oscillators. Several motivations lead to the study of these two models.  See \cite{ito}-\cite{Malekolkalami:2014dca} for more details.

 The paper is organized as follows. In the section \eqref{sec2}, we provide the dynamics of a harmonic oscillator   in  NC spacetime. We  show how this noncommutativity modifies the corresponding metric. In the section \eqref{sec3},  we study the eigen-energies of the Dirac oscillator in the background of cosmic string. The same question is pointed out in the case of the KG  oscillator.     Section \eqref{sec4}   is devoted to   concluding remarks.

\section{Oscillator quantum dynamics on NC spacetime}\label{sec2}
In this section, we study the quantum dynamics on NC spacetime. Using the Hamilton equation of motion of the coordinates system, we derive and solve the corresponding equations of motion. We show, by a novel approach,  how a NC oscillator can be solved in the commutative spacetime, and  how this  may affect the spacetime geometric properties such as the metric tensor. Two cases of noncommutativity are considered. The case, when only coordinates are NC, and the case where both coordinates and momentums are NC.
\subsection{Case of commutative momentum components}
We consider the   spacetime geometry described with the NC coordinates $ x^\mu$ and momentums $ p_\mu$, $\mu=0,1,2,3$, which satisfy the star-commutation relations :
\bea
[ x^\mu, x^\nu]_\star=i\theta^{\mu\nu},\quad [ x^\mu,  p_\nu]_\star=i\delta^\mu_\nu,\quad [ p_\mu,p_\nu]_\star=0,
\eea 
in which,  $\hbar:=1;$  $\star$ denotes the Moyal star product. For $f,g\in C^{\infty}(\mathbb{R}^4\times\mathbb{R}^4)$ we write
\bea
f\star g&=&{\rm \bf m}\big[e^{(\mathcal P_\theta+\mathcal P_{\hbar})}\big],\quad {\rm \bf m}(f\otimes g)=f\cdot g\cr
\mathcal P_\theta&=&\frac{i\theta^{\mu\nu}}{2}\frac{\partial}{\partial x^\mu}\otimes\frac{\partial}{\partial x^\nu},\cr
\mathcal P_{\hbar}&=&\frac{i}{2}\delta^{\mu\nu}\Big(\frac{\partial}{\partial x^\mu} \otimes\frac{\partial}{\partial p^\nu}-\frac{\partial}{\partial p^\mu}\otimes\frac{\partial}{\partial x^\nu}\Big).
\eea
The matrix $\theta^{\mu\nu}$  is chosen to be 
\bea
\theta^{\mu\nu}=\left(\begin{array}{cccc}
0&\theta_0&\theta_0&\theta_0\\
-\theta_0&0&\theta&\theta\\
-\theta_0&-\theta&0&\theta\\
-\theta_0&-\theta&-\theta&0
\end{array}
\right),\quad \theta_0,\,\,\theta\in\mathbb{R}
\eea
For instance, setting $\theta_0=0$  means that the time does not commute with the space coordinates and  plays the role of evolution parameter. For all smooth function   of coordinates and momentums $f(x,p)$,   
 we have the following identities 
\bea\label{transmath}
[x^\mu,f(x,p)]_\star&=&i\theta^{\mu\beta}\frac{\partial f(x,p)}{\partial x^\beta}+i\delta^{\mu\beta}\frac{\partial f(x,p)}{\partial p^\beta},\cr
[p^\mu,f(x,p)]_\star&=&-i\delta^{\mu\beta}\frac{\partial f(x,p)}{\partial x^\beta}.
\eea
It is obvious that the NC coordinates are related to the commutative coordinates by the  followings transformations: 
\bea\label{noncom}
\ x^\mu\rightarrow x_c^\mu-\frac{\theta^{\mu\nu}}{2}p_{\nu,c},\quad  p_\mu\rightarrow
p_{\mu,c},
\eea 
where the commutative variables satisfy the commutation relations $[x_c^\mu,x_c^\nu]_\star=0$ and  $[x_c^\mu,p_{\nu,c}]_\star=i\delta^\mu_\nu$. 
Consider  the Hamiltonian system, with Hamiltonian $H\in C^2(\mathbb{R}^8,\mathbb{R})$,  $(x,p)\in \mathbb{R}^4\times \mathbb{R}^4$. 
The Hamiltonian $H$  does not explicitly depend on the time $x_c^0$. Using the Taylor expansion   we write: 
\bea\label{hamto}
&&H(x,p)=\frac{|p|_c^2}{2}+V(x_c)+\cr
&&\sum_{n=1}^\infty\frac{(-1)^n}{n!}\frac{(\theta p_c)^{j_1}}{2}\frac{(\theta p_c)^{j_2}}{2}
\cdots \frac{(\theta p_c)^{j_n}}{2}\frac{d^nV(x_c)}{
dx_c^{j_1}dx_c^{j_2}\cdots dx_c^{j_n}},\cr
&&(\theta p_c)^j=\theta^{j\ell}p_{\ell,c},\,\, j,\ell=1,2,3.
\eea
The  equations of the dynamics associated to the coordinates and momentums with the Hamiltonian \eqref{hamto} are 
\bea \label{eqm}
\frac{ d x_c^\mu}{dx_c^0}=i[x_c^\mu,H]_\star,\quad \frac{ d p_c^\mu}{dx_c^0}=i[p_c^\mu,H]_\star.
\eea 
For any choice of potential $V(x),$ the above equations lead to a cumbersome  system of nonlinear differential equations, not easily to solve.
In the case where $V(x_c)$ is the harmonic oscillator potential, i.e. $V(x)=\frac{M |x_c|^2}{2}$, the Hamiltonian $H$ takes the form
\bea
H=\frac{1}{2}|p_c|^2
+\frac{M}{2}\Big(|x_c|^2-\theta_{ij}x_{c}^ip_{c}^j+\frac{1}{4}\theta^{ij}\theta_{ik} p_{j,c}p_c^k\Big).
\eea
We  show in the sequel that the corresponding equations of motion can be solved in this particular case.
These  equations of motion   are explicitly given by the following system:
\bea\label{posidess}
\begin{cases}
\dot x_c^1\phantom{-}&= -\frac{1}{2}M\theta x_c^2-\frac{1}{2}M\theta x_c^3-(1+\frac{M\theta^2}{2} )p_c^1\\
&-\frac{1}{4}M\theta^2 p_c^2+\frac{1}{4}M\theta^2p_c^3\\
 \dot x_c^2\phantom{-}&=\frac{1}{2}M\theta x_c^1-\frac{1}{2}M\theta x_c^3-(1+\frac{M\theta^2}{2} )p_c^2\\
&-\frac{1}{4}M\theta^2 p_c^1-\frac{1}{4}M\theta^2 p_c^3\cr
 \dot x_c^3\phantom{-}&=\frac{1}{2}M\theta x_c^1+\frac{1}{2}M\theta x_c^2-(1+\frac{M\theta^2}{2} )p_c^3\\
&+\frac{1}{4}M\theta^2 p_c^1-\frac{1}{4}M\theta^2p_c^2\\
\dot p_c^1\phantom{-}&=M x_c^1-\frac{1}{2}M\theta p_c^2-\frac{1}{2}M\theta p_c^3\\
\dot p_c^2\phantom{-} &=M x_c^2+\frac{1}{2}M\theta p_c^1-\frac{1}{2}M\theta p_c^3\\
 \dot p_c^3\phantom{-}&=M x_c^3+\frac{1}{2}M\theta p_c^1+\frac{1}{2}M\theta p_c^2
\end{cases}
\eea
where  ``dot'' means  the first order derivative with respect to the time $x_c^0:=t$. The system \eqref{posidess}  can be solved, by using  the expansion series method, to yield the general solutions:
\bea\label{euro1}
x^j_c(t)= \sum_{k=0}^\infty \,\Big[a^j_k (\theta t)^k \cos(\sqrt{M}t)+  \,b^j_k (\theta t)^k\sin(\sqrt{M}t)\Big]\\ 
\label{euro2} p^j_c(t)=\sum_{k=0}^\infty \,\Big[c^j_k (\theta t)^k \cos(\sqrt{M}t)+  d^j_k (\theta t)^k\sin(\sqrt{M}t)\Big].
\eea

Note that $p_c^j=M \dot x_c^j$. Then the series $a_k^j$ and $b_k^j$ are related to $c_k^j$ and $d_k^j$ by the recursive relations
\bea
\begin{cases}
c_k^j=(k+1) M\theta a_{k+1}^j+\sqrt{M^3}b_k^j\\ d_k^j=(k+1) M\theta b_{k+1}^j-\sqrt{M^3}a_k^j
\end{cases}
.
\eea
We consider the particular case in which we  assume that there exist the  constants $\alpha_0$ and $\beta_0$ such that 
\bea\label{recu}
\begin{cases}
\alpha_0 b_k^j=(k+1) M\theta a_{k+1}^j+\sqrt{M^3}b_k^j\\
 \beta_0 a_k^j=(k+1) M\theta b_{k+1}^j-\sqrt{M^3}a_k^j
\end{cases}
.
\eea
Then,  the series $a_k^j$ and $b_k^j$ satisfy the two-term recursive relations
\bea
\Big(\beta_0+\sqrt{M^3}\Big)a_k^j-\frac{(k+1)(k+2)M^2\theta^2}{\alpha_0-\sqrt{M^3}}a_{k+2}^j=0\\
\Big(\alpha_0-\sqrt{M^3}\Big)b_k^j-\frac{(k+1)(k+2)M^2\theta^2}{\beta_0+\sqrt{M^3}}b_{k+2}^j=0
\eea
 the solutions of which are of  the form:
\bea
a_{k}^j&=&\Omega_a^j\frac{(-1)^{\frac{k}{2}}}{k!}\Big(\frac{(\sqrt{M^3}-\alpha_0)(\sqrt{M^3}+\beta_0)}{M^2\theta^2}\Big)^{\frac{k}{2}},\\
b_k^j&=&\Omega_b^j\frac{(-1)^{\frac{k}{2}}}{k!}\Big(\frac{(\sqrt{M^3}-\alpha_0)(\sqrt{M^3}+\beta_0)}{M^2\theta^2}\Big)^{\frac{k}{2}},
\eea
where $\Omega_{a,b}^j$ are real constants.
Letting
$
\eta^2=\frac{(\sqrt{M^3}-\alpha_0)(\sqrt{M^3}+\beta_0)}{M^2}
$
 and  $\alpha_0=\beta_0=\wp$ transforms the expressions \eqref{euro1} and \eqref{euro2} into the following form:
\begin{itemize}
\item For $\sqrt{M^3}>\wp$
\bea\label{houk1}
x_c^j(t)&=&\mathcal{R}_{e}\Big[ e^{i t \eta}(\Omega^j_a\cos\sqrt{M}t+\Omega^j_b\sin\sqrt{M}t)\Big]\cr
&=&\cos( t \eta)(\Omega^j_a\cos\sqrt{M}t+\Omega^j_b\sin\sqrt{M}t),\\
 \label{houk2}p_c^j(t)&=&\mathcal{R}_{e}\Big[\wp e^{i t \eta}(\Omega^j_a\cos\sqrt{M}t+\Omega^j_b\sin\sqrt{M}t)\Big]\cr
&=&\wp \cos( t \eta)(\Omega^j_a\cos\sqrt{M}t+\Omega^j_b\sin\sqrt{M}t).
\eea
\item For $\sqrt{M^3}<\wp$
\bea\label{houk3}
x_c^j(t)&=&e^{-t \eta}(\Omega^j_a\cos\sqrt{M}t+\Omega^j_b\sin\sqrt{M}t)
\\
 \label{houk4}p_c^j(t)&=&\wp e^{- t \eta}(\Omega^j_a\cos\sqrt{M}t+\Omega^j_b\sin\sqrt{M}t).
\eea
\end{itemize}
These expressions represent the solution of the noncommutative oscillator in the commutative variables. They do not depend on the deformation parameter $\theta$, and hence are not affected by
 the noncommutativity of the spacetime.

 Let us examine now how these solutions  may modify the geometry. Let ${\rm g}_c$ and ${\rm g}$ are  the metrics of ordinary and NC spacetime, respectively. We assume that  ${\rm g}_{c,\mu\nu}:=\eta_{\mu\nu}=\mbox{diag}(-1,+1,+1,+1)$. Then using \eqref{noncom}, \eqref{houk1} and \eqref{houk2}  we get
\bea
 {\rm g}&=&\eta_{\mu\nu} dx^{\mu}dx^{\nu}\cr
&=&-dt^2+\sum_{j=1}^3\Big(dx_c^j dx_c^j+\frac{\wp^2\theta^{jk}\theta^{jl}}{4}dx_{k,c}dx_{l,c}\Big),
\eea
which can  explicitly be  written  in the matrix form  as:
\bea
({\rm g}^{\mu\nu})=\left(\begin{array}{cccc}
-1&0&0&0\\
0&1+\frac{\wp^2\theta^2}{2}&\frac{\wp^2\theta^2}{4}&-\frac{\wp^2\theta^2}{4}\\
0&\frac{\wp^2\theta^2}{4}&1+\frac{\wp^2\theta^2}{2}&\frac{\wp^2\theta^2}{4}\\
0&-\frac{\wp^2\theta^2}{4}&\frac{\wp^2\theta^2}{4}&1+\frac{\wp^2\theta^2}{2}
\end{array}\right).
\eea
In the diagonal form we get
\bea\label{diagnew}
{\rm g}_d= diag(-1,\lambda^2_1,\lambda^2_2,\lambda^2_3), 
\eea
with  eigenvalues
\bea\label{diagnew1}
\lambda_1^2=1,\quad \lambda^2_2=1+\frac{3\theta^2\wp^2}{4}=\lambda^2_3,
\eea
and eigenvectors 
\bea
&&{\bf u}_{-1}=(1,0,0,0), \quad  {\bf u}_{\lambda_1}=(0,1,-1,1),\cr
&& {\bf u}_{\lambda_2}=(0,-1,0,1), \quad {\bf u}_{\lambda_3}=(0,1,1,0).
\eea
Then the determinant of the metric ${\rm g},$ denoted by $g,$ is
$
det({\rm g})=-\Big(1+\frac{3\theta^2\wp^2}{2}+\frac{9\theta^4\wp^4}{16}\Big).
$
In a compact form, we get:  
\bea\label{newmetrique}
{\rm g}&=&-dt^2+ \sum_{j=1}^3 a_j^2(\theta)(dx_{c}^j)^2, \quad a_j(\theta)=\lambda_j,
\eea
 in which the  parameters $a_j(\theta)=\lambda_j,\, j=1,2,3$ play the role of the scale factors. 
\subsection{Case of noncommutative momentum components}
Here we consider the quantum  spacetime described with the NC coordinates $ x^\mu$ and momentums $ p_\mu$, $\mu=0,1,2,3$, which satisfy the star-commutation relations :
\bea
[ x^\mu, x^\nu]_\star=i\theta^{\mu\nu},\,\, [ x^\mu,  p_\nu]_\star=i\kappa^\mu_\nu,\,\, [ p_\mu,p_\nu]_\star=i\bar{\theta}_{\mu\nu},
\eea 
where the Moyal star product takes the form
\bea
f\star g&=&{\rm\bf m}\big[\exp(\mathcal P_\theta+\mathcal P_{\bar\theta}+\mathcal P_{\hbar})\big]\cr
\mathcal P_{\bar\theta}&=&\frac{i\bar\theta^{\mu\nu}}{2}\frac{\partial}{\partial p^\mu}\otimes\frac{\partial}{\partial p^\nu}.
\eea
The skew symmetric matrix $(\bar \theta)$ is chosen to be
\bea
\bar\theta^{\mu\nu}=\left(\begin{array}{cccc}
0&\bar\theta_0&\bar\theta_0&\bar\theta_0\\
-\bar\theta_0&0&\bar\theta&\bar\theta\\
-\bar\theta_0&-\bar\theta&0&\bar\theta\\
-\bar\theta_0&-\bar\theta&-\bar\theta&0
\end{array}
\right),\quad \bar\theta_0,\,\,\bar\theta\in\mathbb{R}
\eea
For instance,  by choosing $\bar\theta_0=0,$  the NC coordinates are related to the commutative coordinates by the relations:
\bea
x^\mu=x_{c}^\mu-\frac{1}{2}\theta^{\mu\nu}p_{\nu,c},\quad p^\mu=p_{c}^\mu+\frac{1}{2}\bar\theta^{\mu\nu}x_{\nu,c},
\eea
such that the following commutation relations hold:  $[x_c^\mu,x_c^\nu]_\star=0$, $[x_c^\mu,p_{\nu,c}]_\star=i\delta^\mu_\nu$, $[p_c^\mu,p^{\nu}_{c}]_\star=0.$ The tensor $\kappa$ takes the form:
\bea
\kappa^{\mu\nu}=\Big(1+\frac{\theta\bar\theta}{4}\Big)\delta^{\mu\nu}.
\eea
Like \eqref{transmath} we get
\bea
[x^\mu,f(x,p)]_\star&=&i\theta^{\mu\beta}\frac{\partial f(x,p)}{\partial x^\beta}+i\delta^{\mu\beta}\frac{\partial f(x,p)}{\partial p^\beta},\cr
[p^\mu,f(x,p)]_\star&=&\frac{i\bar\theta^{\mu\beta}}{2}\frac{\partial f(x,p)}{\partial p^\beta}-i\delta^{\mu\beta}\frac{\partial f(x,p)}{\partial x^\beta}.
\eea
Then the Hamiltonian of the NC harmonic oscillator can be written as:
\bea
H&=&\frac{1}{2}\Big(|p_c|^2+\bar\theta_{ij}p_c^i x_c^j+\frac{1}{4}\bar\theta^{ij}\bar\theta_{ik} x_{j,c}x_c^k\Big)\cr
&+&\frac{M}{2}\Big(|x_c|^2-\theta_{ij}x_{c}^ip_{c}^j+\frac{1}{4}\theta^{ij}\theta_{ik} p_{j,c}p_c^k\Big).
\eea
This leads to  the following sytem of equations of motion
\bea\label{poside}
\begin{cases}
\dot x_c^1\phantom{-}&= -\frac{1}{2}(M\theta+\bar\theta) x_c^2-\frac{1}{2}(M\theta+\bar\theta) x_c^3\\
&-(1+\frac{M\theta^2}{2} )p_c^1-\frac{1}{4}M\theta^2 p_c^2+\frac{1}{4}M\theta^2p_c^3\\
 \dot x_c^2\phantom{-}&=\frac{1}{2}(M\theta+\bar\theta) x_c^1-\frac{1}{2}(M\theta+\bar\theta) x_c^3\\
&-(1+\frac{M\theta^2}{2} )p_c^2-\frac{1}{4}M\theta^2 p_c^1-\frac{1}{4}M\theta^2 p_c^3\\
 \dot x_c^3\phantom{-}&=\frac{1}{2}(M\theta+\bar\theta) x_c^1+\frac{1}{2}(M\theta+\bar\theta) x_c^2\\
&-(1+\frac{M\theta^2}{2} )p_c^3+\frac{1}{4}M\theta^2 p_c^1-\frac{1}{4}M\theta^2p_c^2\\
\dot p_c^1\phantom{-}&=-\frac{1}{2}(M\theta+\bar\theta) p_c^2-\frac{1}{2}(M\theta+\bar\theta) p_c^3\\
&+(M +\frac{\bar\theta^2}{2})x_c^1+\frac{1}{4}\bar\theta^2 x_c^2-\frac{1}{4}\bar\theta^2x_c^3\\
\dot p_c^2\phantom{-} &=\frac{1}{2}(M\theta+\bar\theta) p_c^1-\frac{1}{2}(M\theta+\bar\theta) p_c^3\\
&+(M +\frac{\bar\theta^2}{2})x_c^2+\frac{1}{4}\bar\theta^2 x_c^1+\frac{1}{4}\bar\theta^2 x_c^3\\
 \dot p_c^3\phantom{-}&=\frac{1}{2}(M\theta+\bar\theta) p_c^1+\frac{1}{2}(M\theta +\bar\theta)p_c^2\\
&+(M +\frac{\bar\theta^2}{2})x_c^2-\frac{1}{4}\bar\theta^2 x_c^1+\frac{1}{4}\bar\theta^2x_c^2
\end{cases}
.
\eea
After some algebra, and exploiting  the expansion series method, we obtain:
\bea\label{euroo3}
x^j_c(t)&=& \sum_{k=0}^\infty \sum_{\ell=0}^\infty\,(\theta t+\bar\theta t)^{k+\ell} \Big[a^j_{k\ell} \cos(\sqrt{M}t)\cr
&&+  \,b^j_{k\ell} \sin(\sqrt{M}t)\Big]\\ 
\label{euroo4} p^j_c(t)&=&\sum_{k=0}^\infty \sum_{\ell=0}^\infty\,(\theta t+\bar\theta t)^{k+\ell} \Big[c^j_{k\ell} \cos(\sqrt{M}t)\cr
&&+  \,d^j_{k\ell} \sin(\sqrt{M}t)\Big].
\eea
Now by setting $p_c^j(t)=M \dot x_c^j(t)$,
 we get the following recursive relations:
\bea
c_{kl}^j=\frac{M(\theta+\bar\theta)(k+\ell+1)}{2}(a^j_{k+1,\ell}+a^j_{k,\ell+1})+\sqrt{M^3}b^j_{k\ell}\cr
d_{kl}^j=\frac{M(\theta+\bar\theta)(k+\ell+1)}{2}(b^j_{k+1,\ell}+b^j_{k,\ell+1})-\sqrt{M^3}a^j_{k\ell}\nonumber
\eea
 $\alpha_0$ and $\beta_0$ are two constants such that $c_{k\ell}^j=\alpha_0b_{k\ell}^j$ and $d_{k\ell}^j=\beta_0a_{k\ell}^j.$ We come to 
\bea\label{palma1}
(\beta_0+\sqrt{M^3})a_{k\ell}^j&=&\frac{M^2(\theta+\bar\theta)^2}{4(\alpha_0-\sqrt{M^3})}(k+\ell+1)(k+\ell+2)\cr
&\times&(a_{k+2,\ell}^j+2a^j_{k+1,\ell+1}+a^j_{k,\ell+2})
\eea
and
\bea\label{palma2}
(\alpha_0-\sqrt{M^3})b_{k\ell}^j&=&\frac{M^2(\theta+\bar\theta)^2}{4(\beta_0+\sqrt{M^3})}(k+\ell+1)(k+\ell+2)\cr
&\times&(b_{k+2,\ell}^j+2b^j_{k+1,\ell+1}+b^j_{k,\ell+2}).
\eea
The recursive relations \eqref{palma1} and \eqref{palma2} can be solved by   setting
$a^j_{k\ell}=a^j_{k+\ell}$ and $b^j_{k\ell}=b^j_{k+\ell}$. We then get the solutions: 
\bea
b_k^j=a_{k}^j=\Omega_{b,a}^j\frac{(-1)^{\frac{k}{2}}}{k!}\Big(\frac{(\sqrt{M^3}-\alpha_0)(\sqrt{M^3}+\beta_0)}{M^2(\theta+\bar\theta)^2}\Big)^{\frac{k}{2}}.
\eea
The $x_c^j(t)$ and $p_c^j(t)$ are given by   \eqref{houk1} ,\eqref{houk2},\eqref{houk3} and  \eqref{houk4}.

In the sequel, we perform some illustrations.

\section{Applications to the relativistic particles}\label{sec3}
\subsection{The Dirac Oscillator}
We derive  equation governing the Dirac oscillator in noncommutative space in the  background of cosmic string. The model is described in the cylinder coordinates with the FLRW  metric 
\bea\label{metric}
ds^2=-dt^{2}+\lambda_{1}^2(t)dr^{2}+\lambda_{2}^2(t)\alpha^{2}r^{2}d\varphi^{2}+\lambda_{3}^2(t)dz^{2},
\eea
$-\infty< (t,z)<\infty$, $r\geq 0$ and $0\leq \varphi\leq 2\pi$. The parameter $\alpha$ is related to the linear mass density $\tilde{M}$ of the string by $\alpha = 1-4\tilde{M}$ and  belongs to   the interval $(0,1],$ corresponding to a deficit angle $\g = 2\pi(1-\alpha)$.  We choose the scale factors $\lambda_1(t),\,\,\lambda_2(t)$ and $\lambda_3(t)$ to be now functions of time and implicitly on $\theta$. The particular case where these three parameters $\lambda_j$ are constant  depending on $\theta$ (see \eqref{newmetrique}) will be discussed hereafter.
 In accordance with the metric \eqref{metric} the tetrad $e_a^\mu(x)$ such that ${\rm g}^{\mu\nu}=e_a^\mu e_b^\nu \eta^{ab}$  is  chosen to be
\bea\label{tetrad}
\big[e^{\mu}_{a}\big] =  \left(\begin{array}{cccc}
1 & 0 & 0& 0\cr
 0 & \frac{\cos\varphi}{\lambda_1} & \frac{\sin\varphi}{\lambda_1} &0\cr
 0 & -\frac{\sin\varphi}{\lambda_2\alpha r} & \frac{\cos\varphi}{\lambda_2\alpha r} & 0\cr
 0 & 0 & 0 & \frac{1}{\lambda_3}
\end{array}\right)
\eea
where the Greek indices is related to the curve space indices and the Latin indices to the Minkowski space indices. Remark that the tetrad \eqref{tetrad} is not uniquely defined. Any tetrad is related to  \eqref{tetrad} by the local  Lorentz transformation $\Lambda_b^a$  as $e_a^\mu(x)=\Lambda_a^b(x) e_b^\mu(x)$.
 The spinor connection is defined by
\bea\label{Ga}
\Gamma_{\mu} &=&  -\frac{1}{8}\omega_{\mu}^{cd}\big[\g_{c},\g_{d}\big],\\ \omega_{\mu}^{ab} 
&=& e_{\nu}^{a}\Gamma^{\nu}_{\mu\sigma}e^{\sigma}_{c}
\eta^{bc}-
\eta^{bc}e^{\nu}_{c}\partial_{\mu}e_{\nu}^{a}.
\eea
 $\g^{a}$ are the  Dirac matrices in Minkowski space and $ \Gamma^{\nu}_{\mu\sigma}$ is the Christoffel symbol. 
We also use the following notations related to the curve cylindrical coordinates: $(\mu,\nu) = (t, r,\varphi, z)$ and $(a,b) = (0, 1, 2,3)$ for the Minkowki space.  Using \eqref{tetrad} we can  show that $\Gamma_\mu = (0,\Gamma_{r},\Gamma_{\varphi}, \Gamma_{z})$, where
\bea\label{gamma}
\Gamma_{r} &=& -\frac{\dot{\lambda}_1}{2}\g_{0}\g_{1}\cos\varphi - \frac{\dot{\lambda}_1}{2}\g_{0}\g_{2} \sin\varphi,\cr
\Gamma_{\varphi} &=&\frac{\dot{\lambda}_2}{2}\alpha r \g_{0}\g_{1}\sin\varphi -\frac{\dot{\lambda}_2}{2}\alpha r \g_{0}\g_{2}\cos\varphi\cr
&&-\frac{1}{2}(1-\frac{\lambda_2}{\lambda_1}\alpha)\g_{1}\g_{2},\cr
 \Gamma_{z} &=& -\frac{\dot{\lambda}_3}{2}\g_{0}\g_{3}.
\eea
In the above relation $\dot \lambda_j=\frac{d \lambda}{dt}$. The Dirac matices are $\g^\mu=e^{\mu}_{\ell}\g^{\ell}$,  explicitly writen  as:
\bea\label{g}
\g^{t}   &=& \g^{0},\cr
\g^{r}  &=& \frac{\g^{1}}{\lambda_1}\cos\varphi + \frac{\g^{2}}{\lambda_1}\sin\varphi,\cr 
\g^{z} &=&  \frac{\g^{3}}{\lambda_3},\cr
\g^{\varphi} &=& -\frac{\g^{1}}{\lambda_2\alpha r}\sin\varphi + \frac{\g^{2}}{\lambda_2\alpha r}\cos\varphi,
\eea
where we  take the standard Dirac matrix to be
$$
\g^{0} =\left( \begin{array}{ccc}
1& 0\cr
0 & -1
\end{array}\right), \,\,\g^{i} =\left( \begin{array}{ccc}
0& \sigma^{i}\cr
-\sigma^{i} & 0
\end{array}\right),\,\, i=1,2,3.
$$
$\sigma^i$ are the Pauli matrices.
The Dirac equation is the Euler-Lagrange equation of motion of the action
\bea
S[\psi,\bar\psi,\Gamma]=\int\,\sqrt{-g}\, d^4x\,\bar\psi\hat{\mathcal M} \psi
\eea
where $g=det\, {\rm g}^{\mu\nu}=-\lambda_1^2\lambda_2^2\lambda_3^2 \alpha^2 r^2$ and 
\bea
\hat{\mathcal M}=i\gamma^\mu(\nabla_{\mu}+\Gamma_{\mu})-i \gamma^{r}\gamma^0M\omega r-M.
\eea
We get
\bea\label{dirac}
\Big[i\g^{\mu}(\nabla_{\mu}+\Gamma_{\mu} )-i \gamma^{r}\gamma^0M\omega r-M\Big]\psi = 0.
\eea
  $M$ is a mass of the  Dirac particle, $\psi$ is a spinor four-components  of the wave function;  $\nabla_\mu$ is
\bea\label{nabla}
\nabla_\mu=h_\mu^{-1}\frac{\partial}{\partial x^\mu}, \quad {\rm g}^{\mu\nu}=h_\mu^{-2}\eta^{\mu\nu}.
\eea
In general $\nabla_\mu$ given in \eqref{nabla} is not a Hermitian operator and  its components do  not commute. So, the following definition will be used:
 \bea\label{gradian} 
\nabla_{\mu}(\cdot) = \vert g\vert^{-1/4}\partial_\mu(\vert g\vert^{1/4}\cdot),
\eea
such that, for ${\rm g}^{\mu\nu}$ defined in \eqref{metric} we find:
\bea\label{gradian} 
\nabla_{0} &=&\frac{1}{2}\Big( \frac{\dot \lambda_1}{\lambda_1} +\frac{\dot \lambda_2}{\lambda_2} +\frac{\dot \lambda_3}{\lambda_3}\Big) + \frac{\partial}{\partial t},\\ \nabla_r&=&\frac{1}{2r} + \frac{\partial}{\partial r},\\
\nabla_{\varphi}&=& \frac{\partial}{\partial \varphi}, \\
 \nabla_{z}&=&\frac{\partial}{\partial z}.
\eea
These expressions mean that the wave equation $\varepsilon_n^0\psi=i\partial_{t}\psi$ is modified as
\bea
\varepsilon_n\psi= i\nabla_0\psi= i\Big[\frac{1}{2}\Big( \frac{\dot \lambda_1}{\lambda_1} +\frac{\dot \lambda_2}{\lambda_2} +\frac{\dot \lambda_3}{\lambda_3}\Big)  + \frac{\partial}{\partial t}\Big]\psi.
\eea
In the case where $\lambda_j,\,\,j=1,2,3,$ are such that  $\dot\lambda_j/\lambda_j$ are constants, the energy spectrum takes the form
\beq
\varepsilon_n=\varepsilon_n^0+\frac{i}{2}\Big( \frac{\dot \lambda_1}{\lambda_1} +\frac{\dot \lambda_2}{\lambda_2} +\frac{\dot \lambda_3}{\lambda_3}\Big) =\varepsilon_n^0+\varepsilon^{\lambda}_n
\eeq

Now let us assume that $\lambda_j,\, j=1,2,3$ are the constants given in  \eqref{diagnew1}, and  
 the particle moves in $(x,y)$ plane. Then, $\varepsilon_n =\varepsilon_n^0$ and $\varepsilon_n^0$ need to be computed. We   use  the variables  separation method in the Dirac equation \eqref{dirac} as follows:
\beq\label{sepa}
\psi(t,r,\varphi,z) =e^{-i\varepsilon^0_n t }\left(\begin{array}{ccc}
\widetilde{\psi}_{a}(r,\varphi)\cr\widetilde{\psi}_{b}(r,\varphi) 
\end{array}\right).
\eeq
The $z$ dependence of the wave function is  removed due to the phase factor of the form $e^{ik z}$, in which $k$ may be vanished. Using \eqref{sepa}
we  come to the two following  differential equations:
\bea\label{Sphia}
&&i\lambda_1\Big(\sigma^{1}\cos\varphi +\sigma^{2}\sin\varphi\Big)\Big(M-\varepsilon_n^{0} \Big)\widetilde{\psi}_{A} \cr
&&+ \Big[\frac{\partial}{\partial r} +M\omega r -\Big(\frac{\lambda_1}{2\alpha\lambda_2}-1\Big)\frac{1}{ r} + \frac{\lambda_1}{\lambda_2} \frac{i\sigma^3}{\alpha r}\frac{\partial}{\partial \varphi} \Big] \widetilde{\psi}_{B}= 0\cr
&&\\
\label{Sphib}
 &&i\lambda_1\Big(\sigma^{1}\cos\varphi +\sigma^2\sin\varphi\Big)\Big(M+\varepsilon_n^0\Big)\widetilde{\psi}_{B} \cr
&&-\Big[\frac{\partial}{\partial r}-M\omega r -\Big(\frac{\lambda_1}{2\alpha\lambda_2}-1\Big)\frac{1}{ r}+ \frac{\lambda_1}{\lambda_2}\frac{i\sigma^3}{\alpha r}\frac{\partial}{\partial \varphi}\Big] \widetilde{\psi}_{A}= 0\cr
&&
\eea

Let us define the  differential operators:
\bea
H_1&=&\Big[\frac{d}{dr}  +M\omega r+\frac{\lambda_1}{\lambda_2 r}\Big(\frac{1}{2\alpha }+\frac{m\sigma^3}{\alpha }+\frac{\lambda_2}{\lambda_1}\Big)\Big]\cr
&\times&\Big[  \frac{d}{dr} -M\omega r-\frac{\lambda_1}{\lambda_2 r}\Big(\frac{1}{2\alpha }+\frac{m\sigma^3}{\alpha }-\frac{\lambda_2}{\lambda_1}\Big)\Big],\\
H_2&=&\Big[\frac{d}{dr}  +M\omega r-\frac{\lambda_1}{\lambda_2 r}\Big(\frac{1}{2\alpha }+\frac{m\sigma^3}{\alpha }-\frac{\lambda_2}{\lambda_1}\Big)\Big]\cr
&\times&\Big[  \frac{d}{dr} -M\omega r+\frac{\lambda_1}{\lambda_2 r}\Big(\frac{1}{2\alpha }+\frac{m\sigma^3}{\alpha }+\frac{\lambda_2}{\lambda_1}\Big)\ \label{OB2}\Big].
\eea 
By setting
\bea
&&\varepsilon_{a} = \lambda_1\big(M-\varepsilon_n^0 ), \,\,\,\varepsilon_{b} = \lambda_1\big(M+\varepsilon_n^0),\\  
&&\widetilde{\psi}_{a}(r,\varphi) = e^{im\varphi}\widetilde{\psi}_{a} (r),\,\,\, \widetilde{\psi}_{b}(r,\varphi) = e^{im\varphi}\widetilde{\psi}_{b} (r)
\eea
we arrive at  the eigenvalue problems 
\bea
H_1 \widetilde{\psi}_{a}=\varepsilon_{a}\varepsilon_{b}\widetilde{\psi}_{a},\quad   H_2\widetilde{\psi}_{b}=\varepsilon_{a}\varepsilon_{b}\widetilde{\psi}_{b}.
\eea
 Now let us recast the spinors  $\widetilde{\psi}_{a}$ and $\widetilde{\psi}_{b}$ as 
\bea
\widetilde{\psi}_{a}=\left(\begin{array}{cc}
\widetilde{\psi}_{a1}\\
\widetilde{\psi}_{a2}
\end{array}\right),\quad \widetilde{\psi}_{b}=\left(\begin{array}{cc}
\widetilde{\psi}_{b1}\\
\widetilde{\psi}_{b2}
\end{array}\right).
\eea
 The functions $\widetilde{\psi}_{ai},i=1,2$ and $\widetilde{\psi}_{bi},i=1,2$ satisfy the well known Laguere polynomial equation:
\bea\label{eqhermite}
\frac{d^{2}\widetilde{\psi}_{l}}{dr^{2}}+\frac{2}{ r}\frac{d\widetilde{\psi}_{l}}{dr} +\Big(\frac{\eta_{2}}{r^{2}}- M^{2}\omega^{2}r^{2}+\eta_{0}\Big)\widetilde{\psi}_{l} (r)= 0,
 \eea
 where $l=a1,\,\,a2  \mbox{ or }
b1, \,\,b2,$ and $\eta_0,\,\,\eta_2$ are two constants  depending on  $l,$  given by
\bea
\begin{cases}
\phantom{-}&
\eta_{0} =  -\varepsilon_a\varepsilon_b -M\omega\Big(1+\frac{1+2m }{\alpha}\frac{\lambda_1}{\lambda_2}\Big)\\
\phantom{-}&\eta_{2} =  -\frac{\lambda_{1}^{2}}{\lambda_{2}^{2}}\frac{1+2m}{2\alpha}\Big(\frac{1+2m}{2\alpha}-\frac{\lambda_2}{\lambda_1}\Big)
\end{cases}\,\, l=a1\\
 \begin{cases}
\phantom{-}&\eta_{0} =   -\varepsilon_a\varepsilon_b-M\omega\Big(1+\frac{1-2m}{\alpha}\frac{\lambda_1}{\lambda_2}\Big)\\
\phantom{-}& \eta_{2} = -\frac{\lambda_{1}^{2}}{\lambda_{2}^{2}}\frac{1-2m}{2\alpha}\Big(\frac{1-2m}{2\alpha}-\frac{\lambda_2}{\lambda_1}\Big)
\end{cases}\,\, l=a2\\
 \begin{cases}
\phantom{-}&\eta_{0} =  -\varepsilon_a\varepsilon_b+M\omega\Big(1   +\frac{ 1+2m }{\alpha}\frac{\lambda_1}{\lambda_2}\Big)\\
 \phantom{-}&\eta_{2} = -\frac{\lambda_{1}^{2}}{\lambda_{2}^{2}}\frac{1 +2m}{2\alpha}\Big(\frac{1+2m}{2\alpha}-\frac{\lambda_2}{\lambda_1}\Big)
\end{cases}\,\, l=b1\\
\begin{cases}
\phantom{-}&
 \eta_{0} = -\varepsilon_a\varepsilon_b+M\omega \Big(1  +\frac{ 1-2m }{\alpha}\frac{\lambda_1}{\lambda_2}\Big)\\
\phantom{-}& \eta_{2} = -\frac{\lambda_{1}^{2}}{\lambda_{2}^{2}}\frac{1 -2m}{2\alpha}\Big(\frac{1-2m}{2\alpha}-\frac{\lambda_2}{\lambda_1}\Big)
\end{cases}\,\, l=b2
\eea
The general solution of  equation \eqref{eqhermite} is given by
\bea\label{so1}
  \widetilde{\psi}_{l}(r)& =& \sqrt{\frac{(M\omega)^{\lambda+\frac{1}{2}}}{\pi \Omega(\lambda)}} e^{-\frac{1}{2}M\omega r^{2}} r^{\lambda -\frac{1}{2}}L_{n}^{\lambda}(M\omega r^{2})
\eea
where
\bea
&& \lambda = \frac{1}{2}\sqrt{1-4\eta_2},\\
&&
 n = \frac{\eta_0 -2M\omega -M\omega\sqrt{1-4\eta_{2}}}{4M\omega}\\
&&\Omega(\lambda)=\int_0^\infty\, e^{-z}z^{\lambda+\beta}L_n^\lambda(z)^2dz,
  \eea
The eigenvalues $\varepsilon_n^0$ giving  the energies of the model become
\begin{itemize}
\item $\mbox{for }\,|m|\leq\frac{1}{2}\Big(\frac{\lambda_2}{\lambda_1}\alpha-1\Big) $
\bea\label{mag1}
\begin{cases}
\varepsilon^0_{n} =\Big[ M^2+\frac{4M\omega}{\lambda_{1}^2}(n+1)\Big]^{\frac{1}{2}}\phantom{-}&\\
\varepsilon^0_{n}= \Big[M^2+\frac{2M\omega}{\lambda_{1}^2}\Big(2n+1-\frac{1\pm2m}{\alpha}\frac{\lambda_1}{\lambda_2}\Big)\Big]^{\frac{1}{2}}\phantom{-}&
\end{cases}, 
\eea 
\item $\mbox{for }\,|m|\geq\frac{1}{2}\Big(\frac{\lambda_2}{\lambda_1}\alpha-1\Big)$
\bea\label{mag2}
\begin{cases}
\varepsilon^0_{n} = \Big[M^2+\frac{2M\omega}{\lambda_{1}^2}\Big(2n+1+\frac{1\pm2m}{\alpha}\frac{\lambda_1}{\lambda_2}\Big)\Big]^{\frac{1}{2}}\phantom{-}&\\
\varepsilon_{n}^0=\Big[ M^2+\frac{4M\omega n}{\lambda_{1}^2}\Big]^{\frac{1}{2}}\phantom{-}&
\end{cases}. 
\eea
\end{itemize}
\begin{remark}
Our analysis highlights  a new degeneracy of the energy spectrum for the different values of the parameter  $m$. This result sheds light on the fact that the spectrum \eqref{mag1} does not exist in the  literature and should be  considered as a new feature caused by the bounds of $m$. We have also shown that $\epsilon_n^0$ depends on the deformation parameter $\theta$  or $\lambda_j$. We write $\varepsilon^0_n=\varepsilon_n^0(\theta)$.  
In the commutative limit, in which  $\lambda_j=1$, or $\theta=0$, the expression \eqref{mag1} is not well defined. In this situation the energy spectrum is given in \eqref{mag2}. This corresponds to the results computed in \cite{ito}-\cite{Boumali:2014vqa} (and references therein).  Note that   the same analysis can be made in the case of non vanishing electromagnetic fields  solved in the commutative case in \cite{Andrade:2014uqa}. 
\end{remark}
%
%
%
%
%
%

\subsection{The Klein-Gordon oscillator}
This section aims at  applying the method used in the section \eqref{sec2} to the KG equation.
Consider a scalar field $\Phi$ written in the cylindrical coordinates as $\Phi =\Phi(t,r,\varphi,z)$.
The KG oscillator is given by the following equation, (see \cite{Mohadesi:2004xa} for more details):
\bea\label{KG}
(\square_{\rm g}-M^2)\Phi=0,
\eea
where the Dalembertian operator in the spacetime  defined with the metric  \eqref{metric} is
\bea
\square_{\rm g}=( \nabla_\mu+\Gamma_\mu+M\omega r_\mu)( \nabla^\mu+\Gamma^\mu-M\omega r^{\mu}).
\eea
The Einstein summation is applied in the cylindrical coordiantes , with $\mu = (t, r, \varphi, z)$ and  $r_\mu = (0, r, 0, 0)$.

Suppose that the eigenvalues $\lambda_j,\,\, j=1,2,3$ defined  in \eqref{metric} are constants as expected in \eqref{diagnew1}. By replacing  the relations \eqref{gamma} and \eqref{gradian} in the KG equation \eqref{KG} we come to the differential  equation
\bea\label{sooom}
\Big[\frac{\partial^2}{\partial t^2}+ O(r,\varphi,z)+M^2\Big]\Phi=0
\eea
where the operator $ O(r,\varphi,z)$ is
\bea
O(r,\varphi,z)&=&-\frac{1}{\lambda_{3}^{2}}\frac{\partial^2}{\partial z^2}-\frac{1}{\lambda_{1}^{2}}\Big(\frac{1}{2r}+\frac{\partial}{\partial r}\Big)^{2}\cr &&-\frac{1}{\lambda_{2}^{2}\alpha^2 r^2}\Big[\frac{\partial}{\partial \varphi}\pm\frac{i}{2}\Big(1-\frac{\lambda_2}{\lambda_1}\alpha\Big)\Big]^{2}\cr &&+\frac{3}{2\lambda_{1}^{2}}M\omega +\frac{M^2\omega^2}{\lambda_{1}^{2}}r^2
\eea
 Equation \eqref{sooom} admits the variable separation  as  $\Phi(t, r, \varphi, z) =e^{-i\varepsilon^0_{n}t +ikz+im\varphi} \tilde{\Phi}(r)$, in which the radial function $\tilde\Phi (r)$ satisfies the following  equation:
\bea
\Big\{\frac{d^2}{dr^2}+\frac{1}{r}\frac{d}{dr}+\frac{\tilde\eta_2}{r^2}-M^2\omega^2 r^2+\tilde\eta_0\Big\}\tilde{\Phi} =0,
\eea
where
\bea
 \tilde\eta_2 &=& \Big[\frac{1}{4}+\frac{\lambda_{1}^{2}}{\lambda_{2}^{2}\alpha^2 }\Big(m\pm\frac{1}{2}\Big(1-\frac{\lambda_2}{\lambda_1}\alpha\Big)\Big)^{2}\Big]\cr
\tilde\eta_0 &=&   -\frac{3}{2}M\omega-\lambda_{1}^{2}\big(M^2-\varepsilon_{n}^{0\,\,2}\big).
\eea
The solution of this equation is given in \eqref{so1}.
The energy  spectrum becomes 
\bea
\varepsilon_{n} &=& \pm\Bigg\{M^2 +\frac{1}{\lambda_{1}^{2}}\Big[4n+\frac{3}{2}M\omega +2+2\sqrt{\tilde\eta_2}\Bigg\}^{\frac{1}{2}}.
\eea
The commutative limit corresponding to $\lambda_j=1, \, j=1,2,3$ is readily obtained.

\section{Concluding remarks}\label{sec4}
In this paper we have investigated the dynamics of the harmonic oscillator in NC spacetime. The  differential equation of motion deduced from this analysis has been solved. The corresponding deformation of the spacetime metric has been given. As application, the Dirac and KG oscillators have been explicitly described  in the background of cosmic string. We have proved that the  Dirac oscillator exibits a new degeneracy of the energy spectrum,  unknown in the  literature. 
The case of time dependent eigenvalues $\lambda_j(t)$ may be examined in the core of forthcoming investigations.

\section{Acknowledgments}
EB, MNH and EEN works are partially supported by the Abdus Salam International Centre for Theoretical Physics (ICTP, Trieste, Italy) through the Office
of External Activities (OEA)-Prj-15. The ICMPA is in partnership with
the Daniel Iagolnitzer Foundation (DIF), France.
DOS research at Max-Planck Institute  is supported by the Alexander von Humboldt Foundation.

\section*{Appendix}
\begin{proof}{ of relation \eqref{euro1} and  \eqref{euro2}}\\
Consider the equation of motion \eqref{posidess}, rewriten  as:
\bea\label{mama2}
\frac{d\bf X}{dt}= U \bf X
\eea
where ${\bf X}=(x_c^1,x_c^2,x_c^3,p_c^1,p_c^2,p_c^3)$ and $U$ is a matrix:
\begingroup\footnotesize
\begin{flalign*}
\left(\begin{array}{cccccc}
0&-\frac{M\theta}{2}&-\frac{M\theta}{2}&-(1+\frac{M\theta^2}{2})&-\frac{M\theta^2}{4}&\frac{M\theta^2}{4}\\
\frac{M\theta}{2}&0&-\frac{M\theta}{2}&-\frac{M\theta^2}{4}&-(1+\frac{M\theta^2}{2})&-\frac{M\theta^2}{4}\\
\frac{M\theta}{2}&\frac{M\theta}{2}&0&\frac{M\theta^2}{4}&-\frac{M\theta^2}{4}&-(1+\frac{M\theta^2}{4})\\
M&0&0&0&-\frac{M\theta}{2}&-\frac{M\theta}{2}\\
0&M&0&\frac{M\theta}{2}&0&-\frac{M\theta}{2}\\
0&0&M&\frac{M\theta}{2}&\frac{M\theta}{2}&0
\end{array}\right),\nonumber
\end{flalign*}
\endgroup
with characteristic polynomial 
\bea
P(\lambda,\theta)=\left(\lambda ^2+M\right) \left(\lambda ^4+M^2+\lambda ^2 M \left(3 \theta ^2 M+2\right)\right).
\eea
The  eigenvalues and the eigenvectors of the above matrix can be simply given. Then the exact solution of the equation \eqref{posidess} can be given analytically.  Because of the size of the solutions which can be found with a computer software, we give here an alternative way to approximate the result as follows :
\begin{itemize}
\item All the eigenvalues  can be separated into $\lambda_+^{(\ell)}$ and $\lambda_-^{(\ell)}$, $\ell=1,2,3,$ such that
\bea
\lambda^{(\ell)}_\pm=\pm i\sqrt{M} \pm u^\ell(\theta,M),\,\,\, 
\eea
where $u^\ell(\theta,t)$ are functions  depending on $\theta$ and $M$.
More precisely, the first order computation gives:
\bea
\label{qq1}\lambda_\pm^{(1)}&=&\pm i\sqrt{M},\\
 \label{qq2}\lambda_{\pm}^{(2)}&=&\pm i\sqrt{M}\pm i\frac{M\theta \sqrt{3}}{2}+O(\theta^2),\\
 \label{qq3}\lambda_{\pm}^{(3)}&=&\pm i\sqrt{M}\mp i\frac{M\theta \sqrt{3}}{2}+O(\theta^2),
\eea
corresponding to  eigenvectors
\bea
{\bf v_1}=\Big(\frac{i}{\sqrt{M}},-\frac{i}{\sqrt{M}},\frac{i}{\sqrt{M}},1,-1,1\Big)
\eea
\bea
{\bf v_2}=\Big( -\frac{i}{\sqrt{M}},\frac{i}{\sqrt{M}},-\frac{i}{\sqrt{M}},1,-1,1\Big)
\eea
\bea
{\bf v_3}&=&\Big(-\frac{\sqrt{M}}{2M}(\sqrt{3}+i), -\frac{\sqrt{M}}{2M}(\sqrt{3}-i),\frac{i}{\sqrt{M}},\\
&&\frac{1}{2}(i\sqrt{3}-1), \frac{1}{2}(i\sqrt{3}+1),1\Big)+{\bf O}(\theta^2)
\eea
\bea
{\bf v_4}&=&\Big(-\frac{\sqrt{M}}{2M}(\sqrt{3}-i), -\frac{\sqrt{M}}{2M}(\sqrt{3}+i),-\frac{i}{\sqrt{M}},\\
&&-\frac{1}{2}(i\sqrt{3}+1), -\frac{1}{2}(i\sqrt{3}-1),1\Big)+{\bf O}(\theta^2)
\eea
\bea
{\bf v_5}&=&\Big(\frac{\sqrt{M}}{2M}(\sqrt{3}-i), \frac{\sqrt{M}}{2M}(\sqrt{3}+i),\frac{i}{\sqrt{M}},\\
&&-\frac{1}{2}(i\sqrt{3}+1), -\frac{1}{2}(i\sqrt{3}-1),1\Big)+{\bf O}(\theta^2)
\eea
\bea
{\bf v_6}&=&\Big(\frac{-\sqrt{M}}{2M}(\sqrt{3}+i), \frac{\sqrt{M}}{2M}(\sqrt{3}-i),-\frac{i}{\sqrt{M}},\\
&&\frac{1}{2}(i\sqrt{3}-1), \frac{1}{2}(i\sqrt{3}+1),1\Big)+{\bf O}(\theta^2)
\eea
The solution of the system \eqref{posidess} at the first order in $\theta$ can simply be  written as: 
\bea
x^j_c=(a^j_1+\theta t a^j_2) \cos\sqrt{M}t+(b^j_1+\theta t b^j_2)\sin\sqrt{M}t\\
p^j_c=(c^j_1+\theta t c^j_2) \cos\sqrt{M}t+(d^j_1+\theta t d^j_2)\sin\sqrt{M}t
\eea
where $a_{1,2}^j, b_{1,2}^j, c_{1,2}^j$ and $d_{1,2}^j$ are real constants, which depend on the mass parameter $M$. The same procedure can be extended to higher order terms of $\theta$ as follows:
\item Suppose that the eigenvalue $\lambda$ are reduced to \eqref{qq1}, \eqref{qq2} and \eqref{qq3}, i.e. $u^\ell(\theta,M)\equiv C^\ell\theta$. 
\bea
C^{\ell}=\left\{\begin{array}{ccc}
0,\,\,&\phantom{-}\ell=0\\
\pm \frac{iM \sqrt{3}}{2},\,\,&\phantom{-}\ell=1\\
\mp \frac{iM \sqrt{3}}{2},\,\,&\phantom{-}\ell=2
\end{array}\right.
\eea
The solution of \eqref{mama2} becomes
\bea
X^j&=&\sum_{\pm} \sum_{\ell}r^j_\ell e^{\pm u^\ell(\theta,M)t}e^{\pm i\sqrt{M}t},\,\,\cr
&=&\sum_{\pm} \sum_{\ell}r^j_\ell e^{\pm C^\ell \theta t}e^{\pm i\sqrt{M}t}
\eea
 The constants $r^j_\ell\in\mathbb{R}$ are the components of  the vectors ${\bf v_j},\,\, j=1,2,\cdots,6$, and depend on $\theta$ and $M$.
Setting 
$
A^j=r_\ell^j(e^{C^\ell\theta t}+e^{-C^\ell\theta t})$ and $
B^j=i r_\ell^j(e^{C^\ell\theta t}-e^{-C^\ell\theta t})
$
yields
\beq
X^j=\Big[A^j\cos(\sqrt{M}t)+  \,B^j \sin(\sqrt{M}t)\Big]
\eeq
Using the Taylor expansion of  the quantities $A^j$ and $B^j$,
the solutions \eqref{euro1} and  \eqref{euro2} are well satisfied.
\end{itemize}
\end{proof}


\begin{thebibliography}{16}


\bibitem{Snyder:1946qz} 
  H.~S.~Snyder,
  ``Quantized space-time,''
  Phys.\ Rev.\  {\bf 71}, 38 (1947).




\bibitem{Connes:1994yd} 
  A.~Connes,
  ``Noncommutative Geometry,''
  ISBN-9780121858605.



\bibitem{Connes:1990qp} 
  A.~Connes and J.~Lott,
  Nucl.\ Phys.\ Proc.\ Suppl.\  {\bf 18B}, 29 (1991).



\bibitem{Harms:2006dv} 
  B.~Harms and O.~Micu,
  J.\ Phys.\ A {\bf 40}, 10337 (2007)
  [hep-th/0610081].

\bibitem{Scholtz:2005vg} 
  F.~G.~Scholtz, B.~Chakraborty, S.~Gangopadhyay and J.~Govaerts,
  J.\ Phys.\ A {\bf 38}, 9849 (2005)
  [cond-mat/0509331 [cond-mat.mes-hall]].

\bibitem{Seiberg:1999vs} 
  N.~Seiberg and E.~Witten,
  JHEP {\bf 9909}, 032 (1999)
  [hep-th/9908142].

\bibitem{Zinn-Justin:2014wva} 
  J.~Zinn-Justin,
  J.\ Statist.\ Phys.\  {\bf 157}, 990 (2014)
  doi:10.1007/s10955-014-1103-y
  [arXiv:1410.1635 [math-ph]].

\bibitem{Ambjorn:2014dma} 
  J.~Ambjørn, S.~Khachatryan and A.~Sedrakyan,
  Phys.\ Rev.\ D {\bf 92}, no. 2, 026002 (2015)
  doi:10.1103/PhysRevD.92.026002
  [arXiv:1407.0076 [hep-th]].


\bibitem{Livine:2009zz} 
  E.~R.~Livine,
  Class.\ Quant.\ Grav.\  {\bf 26}, 195014 (2009)
  doi:10.1088/0264-9381/26/19/195014
  [arXiv:0811.1462 [gr-qc]].


\bibitem{Gouba:2016iar} 
  L.~Gouba,
  Int.\ J.\ Mod.\ Phys.\ A {\bf 31}, no. 19, 1630025 (2016)
  doi:10.1142/S0217751X16300258
  [arXiv:1603.07176 [hep-th]].



\bibitem{ito}
D. Ito, K. Mori, and E. Carrieri, Nuovo Cimento A 51, 1119
(1967).


\bibitem{Mosh}
M. Moshinsky and A. Szczepaniak, J. Phys. A: Math. Gen. 22 (1989).
\bibitem{Andrade:2014uqa} 
  F.~M.~Andrade and E.~O.~Silva,
  Eur.\ Phys.\ J.\ C {\bf 74}, no. 12, 3187 (2014)
  doi:10.1140/epjc/s10052-014-3187-6
  [arXiv:1403.4113 [hep-th]].


\bibitem{Szabo:2001kg} 
  R.~J.~Szabo,
  Phys.\ Rept.\  {\bf 378}, 207 (2003)
  doi:10.1016/S0370-1573(03)00059-0
  [hep-th/0109162].
%

\bibitem{Carvalho:2011krd} 
  J.~Carvalho, C.~Furtado and F.~Moraes,
  Phys.\ Rev.\ A {\bf 84}, no. 3, 032109 (2011).
  doi:10.1103/PhysRevA.84.032109



\bibitem{deSousa:2015bvy} 
  M.~S.~Maior de Sousa, R.~F.~Ribeiro and E.~R.~Bezerra de Mello,
  Phys.\ Rev.\ D {\bf 93}, no. 4, 043545 (2016)
  doi:10.1103/PhysRevD.93.043545
  [arXiv:1511.02745 [hep-th]].

\bibitem{Akcay:2016wgl} 
  H.~Akcay and R.~Sever,
  Eur.\ Phys.\ J.\ Plus {\bf 131}, no. 7, 225 (2016).
  doi:10.1140/epjp/i2016-16225-1



\bibitem{Boumali:2014vqa} 
  A.~Boumali and H.~Hassanabadi,
  Can.\ J.\ Phys.\  {\bf 93}, no. 5, 542 (2015).
  doi:10.1139/cjp-2014-0276




\bibitem{Hassanabadi:2013oma} 
  H.~Hassanabadi, S.~S.~Hosseini and S.~Zarrinkamar,
  Chin.\ Phys.\ C {\bf 38}, 063104 (2014).
  doi:10.1088/1674-1137/38/6/063104

\bibitem{Mandal:2012wp} 
  B.~P.~Mandal and S.~K.~Rai,
  Phys.\ Lett.\ A {\bf 376}, 2467 (2012)
  doi:10.1016/j.physleta.2012.07.001
  [arXiv:1203.2714 [hep-th]].

\bibitem{Kukulin:1991mg} 
  V.~I.~Kukulin, G.~Loyola and M.~Moshinsky,
  In *College Park 1991, Hadron'91* 293-301

\bibitem{Mohadesi:2004xa} 
  M.~Mohadesi and B.~Mirza,
  Commun.\ Theor.\ Phys.\  {\bf 42}, 664 (2004)
  [hep-th/0412122].





\bibitem{Silva:2015ysa} 
  E.~O.~Silva, S.~C.~Ulhoa, F.~M.~Andrade, C.~Filgueiras and R.~G.~G.~Amorim,
  Annals Phys.\  {\bf 362}, 739 (2015).
  doi:10.1016/j.aop.2015.09.011


\bibitem{Malekolkalami:2014dca} 
  B.~Malekolkalami, K.~Atazadeh and B.~Vakili,
  Phys.\ Lett.\ B {\bf 739}, 400 (2014)
  doi:10.1016/j.physletb.2014.11.003
  [arXiv:1411.3623 [gr-qc]].

\end{thebibliography}
\end{document}